\documentclass[twocolumn,showpacs,preprintnumbers,amsmath,amssymb]{revtex4}
\usepackage{graphicx}
\usepackage{dcolumn}
\usepackage{bm}

\begin{document}


\title{Entanglement, quantum phase transition and scaling in XXZ chain}

\author{Shi-Jian Gu$^{1,2}$}
\altaffiliation{Email: sjgu@zimp.zju.edu.cn\\URL:
http://zimp.zju.edu.cn/$\widetilde{\;\,}$sjgu/}
\author{Hai-Qing Lin$^1$} \author{You-Quan Li$^2$}
\affiliation{$^1$Department of Physics, The Chinese University of
Hong Kong, Hong Kong, China} \affiliation{$^2$Zhejiang Institute
of Modern Physics, Zhejiang University, Hangzhou 310027, P. R.
China}

\begin{abstract}
Motivated by recent development in quantum entanglement, we study
relations among concurrence $C$, SU$_q$(2) algebra, quantum phase
transition and correlation length at the zero temperature for the
XXZ chain. We find that at the SU(2) point, the ground state
possess the maximum concurrence. When the anisotropic parameter
$\Delta$ is deformed, however, its value decreases. Its dependence
on $\Delta$ scales as $C=C_0-C_1(\Delta-1)^2$ in the XY metallic
phase and near the critical point (i.e. $1<\Delta<1.3$) of the
Ising-like insulating phase. We also study the dependence of $C$
on the correlation length $\xi$, and show that it satisfies
$C=C_0-1/2\xi$ near the critical point. For different size of the
system, we show that there exists a universal scaling function of
$C$ with respect to the correlation length $\xi$.
\end{abstract}

\maketitle

Quantum entanglement, as one of the most intriguing feature of
quantum theory, has been a subject of much study in recent years,
mostly because its nonlocal connotation\cite{ABinstein35} is
regarded as a valuable resource in quantum communication and
information processing\cite{SeeForExample,MANielsenb}. For
example, an entangled state, such as a singlet state
$\frac{1}{\sqrt{2}}(|\uparrow\downarrow\rangle-|\downarrow\uparrow\rangle)$,
can be used for the realization of
teleportation\cite{CHBennett93}. On the other hand, as with other
resources, such as free energy and information, one would like to
know how it can be quantified and controlled. For the first
problem, much efforts have been devoted to develop a quantitative
theory of entanglement, include entanglement of
formation\cite{PRungta2001,CHBennett96,WKWootters98,VCoffman2000},
which is regarded as its basic measure. For the second problem,
many
authors\cite{KMOConnor2001,PZandardi2000,LFSantos2003,XWang2001PLA,XWang2002PLA,XWangPRA2001,GLagmago2002,TJOsbornee,
AOsterloh2002} tried to built bridge between quantum entanglement
and physical models by investigating their entanglement in both
the ground state\cite{KMOConnor2001,TJOsbornee} and thermal
state\cite{XWangPRA2001,GLagmago2002}.

Very recently, the intriguing issue of the relation between
entanglement and quantum phase transition have been
addressed\cite{AOsterloh2002,GVidal2003}. For a spin-1/2
ferromagnetic chain, Osterloh {\it et. al. }, reported that the
entanglement shows scaling behavior in the vicinity of quantum
phase transition point \cite{SSachdevb} as induced by a transverse
magnetic field. Vidal {\it et. al.} tried to establish a
connection between quantum information and condensed matter theory
by studying the behavior of critical entanglement in spin systems.
So it is believed that the entanglement of the ground state, like
the conductivity in the Mott-insulator transition\cite{FGebbhardb}
and quantum Hall effect, and magnetization in the
external-field-induced phase transition, is also plays a crucial
role to the understanding of quantum phase transition. On the
other hand, group theory as well as symmetry of the system are
parts of the foundation of quantum
mechanics\cite{WGreinerb,CKasselb}, the knowledge of its presence
often makes it easy to understand the physics. Thus the study of
entanglement at the ground state and its relation to the group
theory will not only have a contribution to experimental
realization, but also enrichs our physical intuition of quantum
theory.

The main focus of the present paper is to study the properties of
ground state concurrence of an antiferromagnetic XXZ chain.
We show that the competition between quantum fluctuation and
ordering will lead to maximum value of concurrence at the
isotropic point. This observation could also be clarified from the point
view of $q$-deformation theory. The concurrence's dependence on
anisotropic parameter $\Delta$ is presented both numericallly
and analytically. The relation of the concurrence to the correlation
length $\xi$ in the Ising-like insulating phase, as well as the scaling
behavior around the critical point $\Delta=1$ where the
Metal-insulator quantum phase transition occurs, are also
discussed. Thus our result not only manifest interesting physical
phenomenon, but also establish non-trivial connection between the
quantities in quantum information theory and critical phenomenon,
correlation length in condensed matter physics and quantum group
theory\cite{CKasselb}.

The Hamiltonian of the XXZ chain with periodic
boundary conditions reads
\begin{eqnarray}
&& H(\Delta)=\sum_l^N [\sigma_l^x \sigma_{l+1}^x+\sigma_l^y
\sigma_{l+1}^y+\Delta \sigma_l^z \sigma_{l+1}^z],\nonumber \\
&&\sigma_{N+1}=\sigma_1, \label{eq:Hamiltonian}
\end{eqnarray}
where $N$ is the number of sites, $\sigma^\alpha\; (\alpha=x,y,z)$
are Pauli matrices, and $\Delta$ is a dimensionless parameter characterizing
 anisotropic interaction. The
Hamiltonian is invariant under translation, therefore, the
entanglement between arbitrary two neighbor sites is a uniform
function of site index. At $\Delta=1$, Eq. (\ref{eq:Hamiltonian})
has SU(2) symmetry. While $\Delta\neq 1$, it becomes $q$-deformed
SU(2) algebra with $\Delta=(q+q^{-1})/2$. Together with the $Z^2$
symmetry, we can have $[H, S^z]=0$, which result in that the
reduced density matrix $\rho_{l(l+1)}$ of two neighbor sites is of
the form\cite{KMOConnor2001}
\begin{eqnarray}
\rho_{l(l+1)}=\begin{pmatrix}
  u^+ & 0 & 0 & 0 \\
  0 & w_1 & z & 0 \\
  0 & z^* & w_2 & 0 \\
  0 & 0 & 0 & u^-
\end{pmatrix}
\end{eqnarray}
in the standard basis $|\uparrow\uparrow\rangle$,
$|\uparrow\downarrow\rangle$, $|\downarrow\uparrow\rangle$,
$|\downarrow\downarrow\rangle$. Since the energy of a single pair
in the system is $E/N={\rm tr}[\rho_{l(l+1)}H_l]$, where $H_l$ is
the part of Hamiltonian between site $l$ and $l+1$, due to the
translational invariance. Considering the definition of
entanglement, we can easily obtain that the concurrence of XXZ
chain can be calculated as\cite{XWang2002PLA,UGlaser0305108}
\begin{eqnarray}
C=\frac{1}{2}\max[0, |E/N-\Delta
G_{l(l+1)}^{zz}|-G_{l(l+1)}^{zz}-1]. \label{eq:condcdd}
\end{eqnarray}
where $G_{l(l+1)}^{zz}$ is the correlation function. So we not only need to
know the energy of the system, but also the behavior of
correlation function.

It is well known that the present model can be exactly solved by
Quantum Inverse Method \cite{HABethe31,MTakahashib}, and its
energy spectra are determined by a set of spin rapidities
$\lambda_1, \lambda_2,\dots, \lambda_M$, which describe the
kinetic behavior of a state with $M$ down spins. They are the
solution of Bethe-ansatz equation
\begin{eqnarray}
\left(\frac{\sinh\gamma(\lambda_j+i)}
           {\sinh\gamma(\lambda_j-i)}\right)^N
           =\prod_{l\neq j}^{M}
   \frac{\sinh\gamma(\lambda_j-\lambda_l + 2i)}
        {\sinh\gamma(\lambda_j-\lambda_l - 2i)}
\label{app_eq:BAE}
\end{eqnarray}
where the parameter $\gamma$ arises from the anisotropic scale
$\Delta$, i.e., $\Delta=\cos 2\gamma$. The regime $0<\Delta<1$ is
characterized by real positive $\gamma$ while the regime
$1<\Delta$ by pure imaginary $\gamma$ with positive imaginary
part. When $\gamma\rightarrow 0$, the above secular equations
reduce to the well known one for isotropic Heisenberg model.

Taking the logarithm of the above equation, we can have a set of
transcendental equations for $\{\lambda_j\}$, in which the energy
level is determined by a set of quantum number $\{I_j\}$. For the
ground state, $\{I_j\}$ are consecutive integer or
half-odd-integer centering around zero, and $M=N/2$. Then the
ground state energy of the system can be calculated either by
solving the Bethe ansatz equations numerically for finite size
system, or by solving integral equation of density function of
$\lambda$ in the thermodynamic limit. Once the $\Delta$ dependent
eigenenergy $E(\Delta)$ is obtained, the correlation function is
simply the first derivative of $E(\Delta)/N$ with respect to
$\Delta$.

\begin{figure}
\includegraphics[width=7cm]{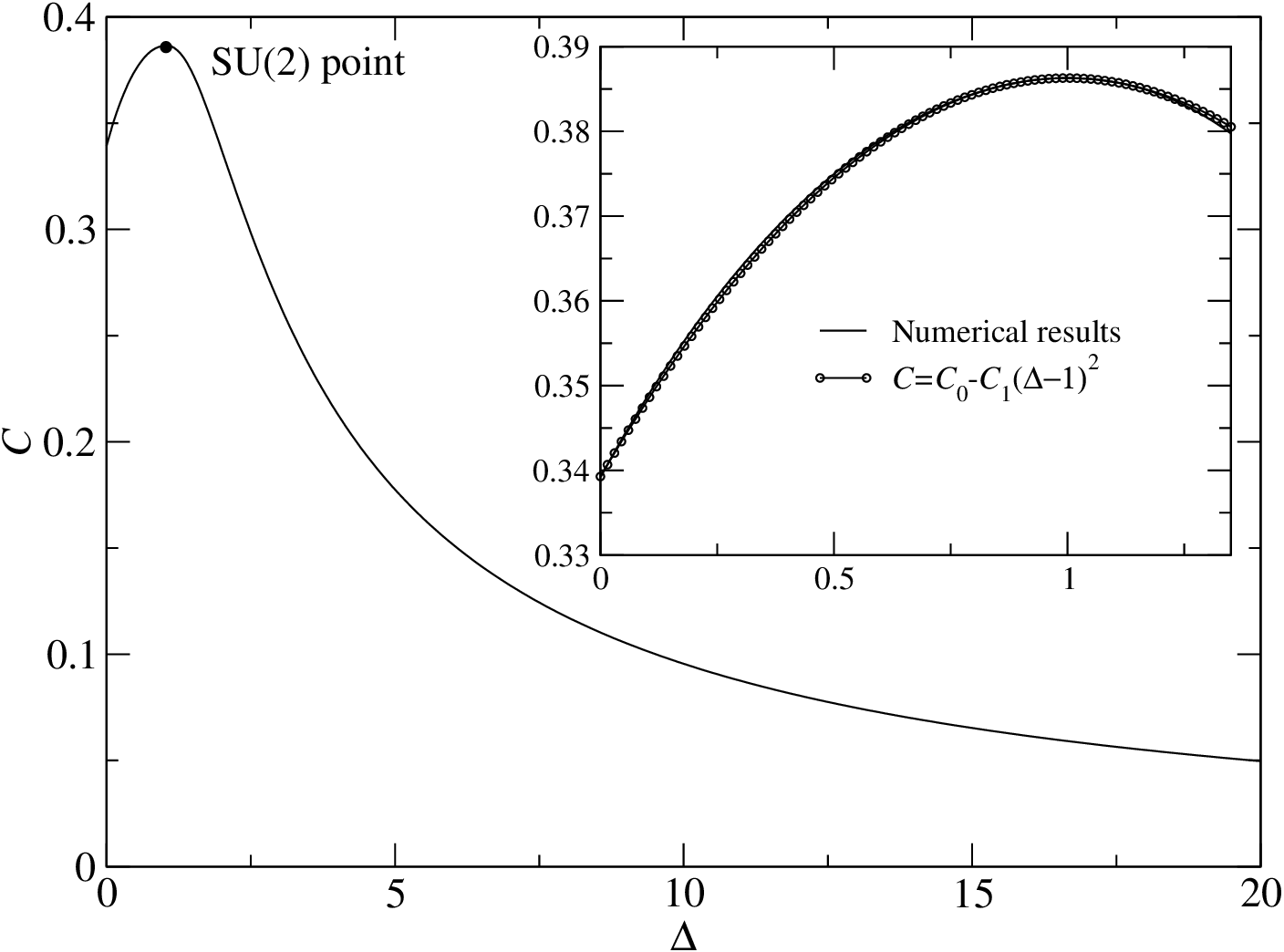}
\caption{\label{figure_concur} Representation of concurrence $C$
as a function of $\Delta$, obtained by solving three sets of Bethe
ansatz equations of $N=1280$ sites system numerically. It is clear
that the concurrence reach is maximum at the critical point
$\Delta=1$.\\}
\end{figure}

For the XXZ model, there exist two different phases at the ground
state, i.e., metallic phase: $0<\Delta\leq 1$ and insulating
phase: $\Delta>1$, which is resulted from that the former is
gapless while the later is gapful. The critical point of quantum
phase transition locates at the isotropic point $\Delta=1$ at
which the concurrence is just a simple function of ground state
energy per sites, i.e. 0.386. If we regard $\sigma^z$ as a
`coordinate', then the first two terms in Eq.
(\ref{eq:Hamiltonian}) represents the `kinetic' energy causing the
quantum fluctuations of $\sigma^z$, and the last term represents
the `potential energy' that causes the ordering of $\sigma^z$. In
the Ising limit $\Delta\rightarrow\infty$, the ground state has
the N\'{e}el long-range-order, which results in that the
concurrence is zero. When $\Delta$ becomes smaller but still large
than 1, the quantum fluctuation plays more and more important
role, then the N\'{e}el state is no longer an eigenstate of the
Hamiltonian. This fluctuation between two neighboring sites
enhances the value of off-diagonal term $z$ in their reduced
density matrix $\rho_{l(l+1)}$, then the entanglement becomes
larger and larger.
On the other hand, at the free particle(XX) limit where
$\Delta=0$, the spin-flip term dominates the system completely,
and all spins flip freely on lattice sites. For a certain site
$j$, the probability of spin up and down is the same, regardless
the spin state of its neighbor. Thus the state
$|\uparrow\uparrow\rangle$ will not lower the energy, but share
the same probability with $|\downarrow\uparrow\rangle$ or
$|\uparrow\downarrow\rangle$. This phenomenon will result in a
relatively large $u^+$ or $u^-$ in the reduced density matrix of
two neighbor sites, as well as a relative smaller $C$. On the
contrary, once the anisotropic interaction $1>\Delta>0$ is turned
on, the value of $u^+$ and $u^-$ is lowered. So the concurrence is
enhanced. Hence the competition of quantum fluctuation and
ordering must results in a maximum concurrence at a certain point.
Comparing with the origin of metal-insulator transition in the
present model, which also arise from the competition of
fluctuation and ordering, it is natural to infer that the point we
want here is just the isotropic point, i.e. $\Delta=1$, as
illustrated in the Fig. \ref{figure_concur}. This case is very
similar to the formation of Kondo effect, in which the competition
between spin singlet formation and thermal conductivity leads to a
minimum conductivity at the Kondo temperature. The idea can also
be applied to the entanglement of arbitrary two sites, such as the
concurrence $C_{lm}$ between site $l$ and $m$. Only when the
competition between their interaction and fluctuation reaches a
counterbalance, the concurrence $C_{lm}$ reaches its maximum.

From the quantum group theory point of view, at $\Delta=1$ point,
the ground state is SU(2) singlet in which the two neighboring
sites try to form antisymmetric pair, as
$(|\uparrow\downarrow\rangle-|\downarrow\uparrow\rangle)/\sqrt{2}$.
In the $q$-deformed region, the Hamiltonian (\ref{eq:Hamiltonian})
can be rewritten in terms of Temperly-Lieb operators
\begin{eqnarray}
H=N\Delta + 2\sum_{j}^N T_{j,j+1},\label{eq:qpHamiltoaian}
\end{eqnarray}
where $T_{j,j+1}=\{-q^{-1}, 1;\;1, -q\}$ in the basis
$|\uparrow\downarrow\rangle, |\downarrow\uparrow\rangle$. Define
$q$-deformed antisymmetric state
$|\phi_q\rangle=(|\uparrow\downarrow\rangle-q|\downarrow\uparrow\rangle)/\sqrt{1+q^2}$,
then the operator $T_{j,j+1}$ can be expressed as
$T_{j,j+1}=-\frac{\Delta}{2}|\phi_q\rangle\langle\phi_q|$. If
$\Delta>1$,
 the lowest energy state favor the
formation of $q$-deformed antisymmetric state between two
neighboring sites\cite{JAbbottphdthesis}, unlike the case of
$\Delta=1$ where it favors antisymmetric state, which obviously
leads to the decrease of concurrence between two neighboring
sites. When the deformation parameter $q$ becomes very large, it
tends to the N\'eel state. On the other hand, the $|\phi_q\rangle$
breaks the local translational invariance, from the point view of
spinless fermions model, the formation of $|\phi_q\rangle$
develops charge-density-state (CDW) at the ground state, which is
gapped and low symmetric.

We show the concurrence as a function of $\Delta$ in Fig.
\ref{figure_concur}, which is obtained by solving both the Bethe
ansatz equations for 1280 sites system numerically, and the
integral equation for infinite length system (We obtained the same
result). As we expect, the ground state at the isotropic point
possesses the maximum concurrence. Thus symmetry of the
Hamiltonian plays a central role in determining the concurrence of
its ground state.  And the trend of curve can be easily understood
based on the above argument. On the other hand, a challenge and
non-trial problem is to quantify the concurrence around the
critical point. In the XY metallic phase and near the critical
point (i.e. $1<\Delta<1.3$) of the Ising-like insulating phase, it
is amazing that $C$ can be described by
\begin{eqnarray}
C=C_0-C_1(\Delta-1)^2, \label{eq:con_XYphase}
\end{eqnarray}
very well, where
\begin{eqnarray}
C_0 &=& 2\ln 2-1 \simeq 0.386,\nonumber\\
C_1 &=& 2\ln 2-\frac{1}{2}-\frac{2}{\pi}-\frac{2}{\pi^2}\simeq
0.047,
\end{eqnarray}
as illustrated in the inset of Fig. \ref{figure_concur}. Hence
around $\Delta=1$, the critical exponents of the anisotropic term
is 2. As we know, the present model can be transformed into
spinless fermions model by Jordan-Wigner transformation. For the
free particle case, it is easy to obtain that the ground state
energy and the correlation function $G^{zz}$ are $4/\pi$ and
$4/\pi^2$ respectively.
In the
large $\Delta$ limit, we find the concurrence scales like
$C\propto 1/\Delta$. One can also express $C$ in terms of
deformation factor $q$ via the relation
$q=\Delta\pm\sqrt{\Delta^2-1}$. It has the form
\begin{eqnarray}
C=C_0-\frac{C_1}{4}(q^{1/2}-q^{-1/2})^4.
\end{eqnarray}
around the critical point. In XY metallic phase, it we define
$q=e^{i\phi}$, it becomes
\begin{eqnarray}
C=C_0-4C_1\sin^4\frac{\phi}{2}.
\end{eqnarray}

\begin{figure}
\includegraphics[width=7cm]{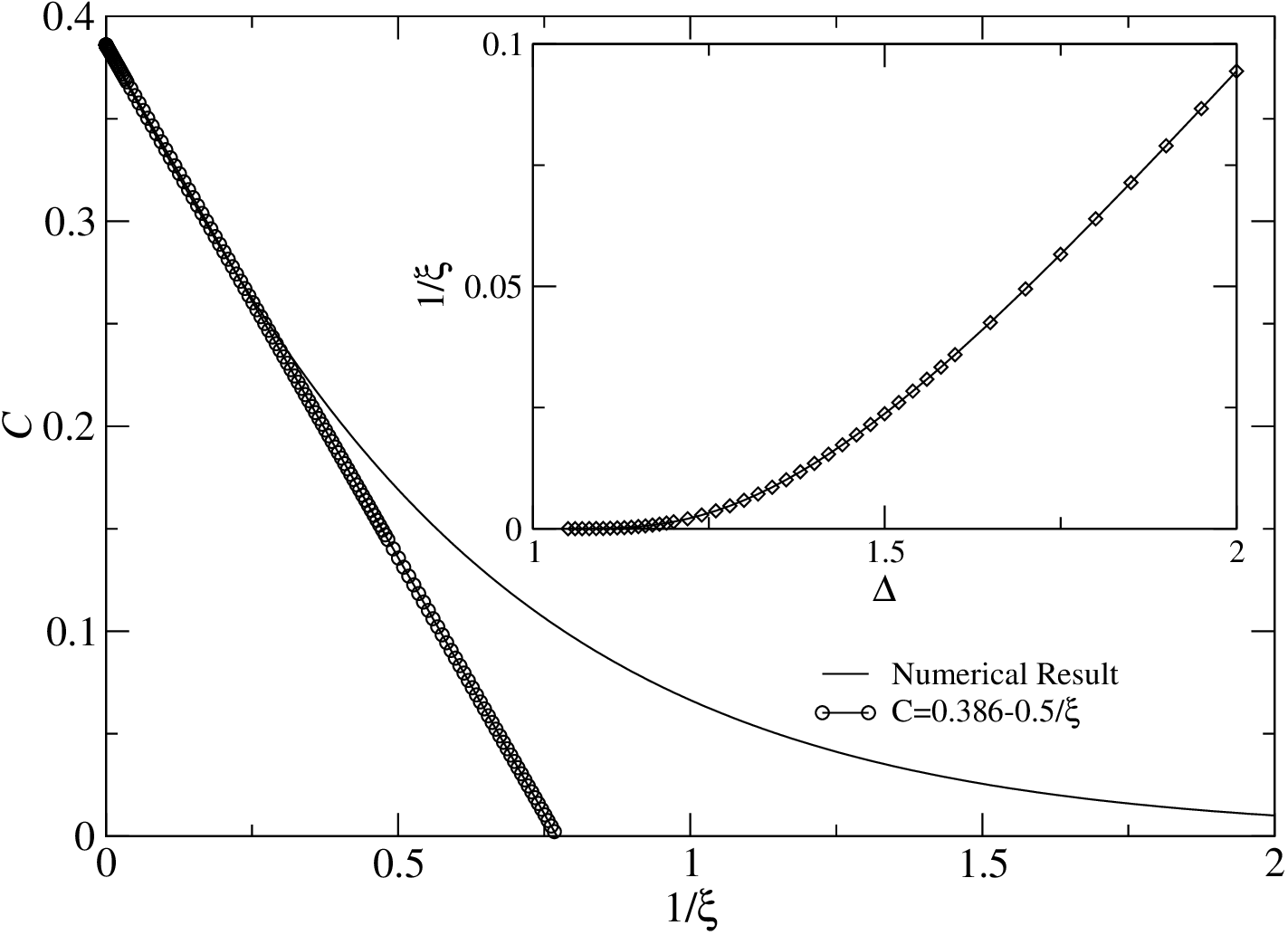}
\caption{\label{figure_crrcon} Representation of concurrence $C$
as a function of $1/\xi$. Here $\xi$ is in unit of lattice constant.}
\end{figure}

\begin{figure}
\includegraphics[width=7cm]{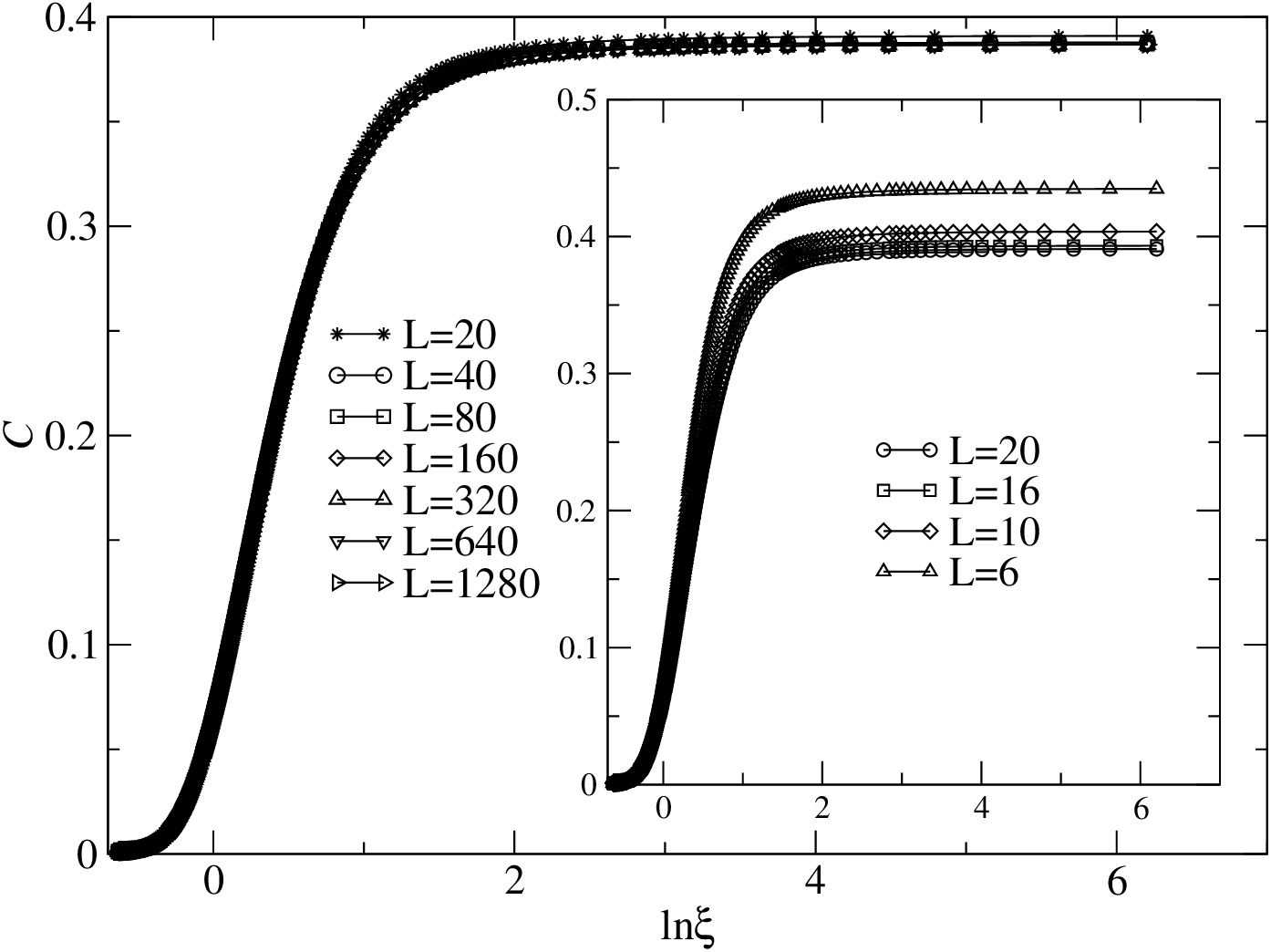}
\caption{\label{figure_scale} Representation of concurrence $C$ as
a function of $\xi$ for different system sizes.}
\end{figure}

Now we study the scaling behavior in the Ising-like insulating phase by
considering the correlation length. Though the scaling study of
metal-insulator transition based on the analysis of spin stiffness
has proposed recently\cite{SJGu2002}, and though everyone believe
there must exist some relation between correlation and
entanglement, the scaling of concurrence, and its dependence on
the correlation length still remains an open and interesting
problem. By analyzing the finite chain system, one can obtain the
correlation length as a function of $\Delta$ in an easy
way\cite{SJGu2002,RJBaxterb}. It has the form
\begin{eqnarray}
1/\xi=\gamma+\sum_{n=1}^\infty \frac{(-1)^n}{n}\tanh(2n\gamma),
\end{eqnarray}
In the $\Delta\rightarrow 1$ limit, it has a good approximation,
as $1/\xi\propto (\Delta-1)^2$. Clearly, the correlation length is
independent of system size, its behavior is shown in the inset of
Fig. \ref{figure_crrcon}. The dependence of $C$ on $\xi$ is
represented in Fig. \ref{figure_crrcon}, in which the solid line
is obtained by solving Bethe ansatz equations for 1280 sites
system numerically. For the value of $\xi$ bigger than $4$, i.e.
$1/\xi<0.25$, there exist a simple relation between $C$ and $\xi$,
which scales
\begin{eqnarray}
C=C_0-\frac{1}{2\xi}. \label{eq:con_crr}
\end{eqnarray}
The above equation imply that the concurrence of does not have a
long-range effect, in another way, we can say that a smallish
system, such as $N=20$, can well describe the behavior of
concurrence of large system, as illustrated in Fig.
\ref{figure_scale}. Compared with the scaling of spin
stiffness\cite{SJGu2002}, the present one is more perfect, that is
the concurrence is almost independent of the system size when
$L>10$. So we can conclude that for finite size system, there
exist a scaling function, which is independent of $L$ and scales
like Eq. (\ref{eq:con_crr}) in large $\xi$ limit. Only when
$L<10$, the finite size effect becomes very clear (See the inset
of Fig. \ref{figure_scale}). Moreover, for small system,
concurrence in even number sites and odd one is different. The
former is usually larger than the later due to the frustration
effect happens in odd sites system with periodic boundary
condition. For example, for 3 sites system, the two singlet
formations between sites 1, 2 and between sites 2, 3 breaks
singlet formation of sites 3, 1.  When $L$ becomes large, this
effects can be neglected and the concurrence in two case is the
same.

In summary, we have investigated the ground state concurrence of
the XXZ chain. We pointed out that the competition between quantum
fluctuation and N\'{e}el ordering will lead to a maximum value of
concurrence at the isotropic Heisenberg point. Based on the Bethe
ansatz solution, we exactly obtained the dependence of $C$ on the
parameter $\Delta$ in a wide range around the critical point, and
obtained numerical result in the whole range.  We established the
relation between the concurrence and deformation factor $q$ of
quantum group in the Ising-like insulating phase. It is now clear
that $q$-deformed permutation generator favors the formation of a
deformed ground state, which has a relatively smaller concurrence.
Moreover, the relation between the concurrence and the correlation
length was studied both numerically and analytically. We found
that there exists a universal scaling behavior for finite(not
small) size system, and it satisfies a simple relation $C\propto
1/2\xi$ in the region close to the critic point.

This work was supported in part by the Earmarked Grant for
Research from the Research Grants Council (RGC) of the HKSAR,
China (Project CUHK 4246/01P) and by NSFC No. 90103022 \&
10225419. S. J. Gu is grateful for the hospitality of the Physics
Department at CUHK. We thank X. Wang and H. Q. Zhou for helpful
discussions.

\end{document}